\documentclass{PoS}

\usepackage{slashbox}
\usepackage{epsfig}
\usepackage{wrapfig}

\newcommand\bef{\begin{figure}}
\newcommand\eef[1]{\label{fg:#1}\end{figure}}
\newcommand\besf{\begin{subfigure}}
\newcommand\eesf[1]{\label{sfg:#1}\end{subfigure}}
\newcommand\beq{\begin{equation}}
\newcommand\eeq[1]{\label{#1}\end{equation}}
\newcommand\beqa{\begin{eqnarray}}
\newcommand\eeqa[1]{\label{#1}\end{eqnarray}}
\newcommand\bet{\begin{table}}
\newcommand\eet[1]{\label{tb:#1}\end{table}}
\newcommand\best{\begin{subtable}}
\newcommand\eest[1]{\label{stb:#1}\end{subtable}}
\newcommand\betb{\begin{center}\begin{tabular}}
\newcommand\eetb{\end{tabular}\end{center}}
\newcommand\beit{\begin{itemize}}
\newcommand\eeit{\end{itemize}}

\newcommand\fgn[1]{Figure \ref{fg:#1}}

\newcommand\tbn[1]{Table \ref{tb:#1}}

\newcommand\incfig[2]{\includegraphics[scale=#1]{#2}}

\title{Spectroscopy of doubly and triply-charmed baryons from lattice QCD}

\ShortTitle{Spectroscopy of doubly and triply-charmed baryons from lattice QCD}

\author{\speaker{M. Padmanath} \\%
        Tata Institute of Fundamental Research, Mumbai\\
        E-mail: \email{padmanath@theory.tifr.res.in}}

\author{Robert\ G.\ Edwards\\
        Jefferson Laboratory, VA\\
        E-mail: \email{edwards@jlab.org}}

\author{Nilmani\ Mathur\\
        Tata Institute of Fundamental Research, Mumbai\\
        E-mail: \email{nilmani@theory.tifr.res.in}}

\author{Michael\ Peardon\\
        Trinity College, Dublin\\
        E-mail: \email{mjp@maths.tcd.ie}}

\abstract{We present the ground and excited state spectra of doubly and triply-charmed 
baryons by using lattice QCD with dynamical clover fermions. A large set of baryonic 
operators that respect the symmetries of the lattice and are obtained after subduction 
from their continuum analogues are utilized. Using novel computational techniques 
correlation functions of these operators are generated and the variational method is 
exploited to extract excited states. The lattice spectra that we obtain have baryonic 
states with well-defined total spins up to 7/2 and the low lying states remarkably 
resemble the expectations of quantum numbers from SU(6) $\otimes$ O(3) symmetry. Various 
energy splittings between the extracted states, including splittings due to hyperfine 
as well as spin-orbit coupling, are considered and those are also compared against 
similar energy splittings at other quark masses. Using those splittings for doubly-charmed 
baryons, and taking input of experimental $B_c$ meson mass, we predict the mass 
splittings of $B^*_c - B_c$ to be about 80 $\pm$ 8 MeV and $m_{\Omega_{ccb}} = 8050\pm10$ MeV.}

\FullConference{31$^{st}$ International Symposium on Lattice Field Theory LATTICE 2013\\
                 July 29 - August 3, 2013\\
                 Mainz, Germany}

\begin{document}

\section{Introduction}

\vspace{-0.4cm}
The recent discovery of numerous hadrons at various particle colliders, like Belle, BaBar,
CDF, LHCb, BECIII etc., has brought a resurgence of interest in the heavy hadron spectroscopy 
\cite{PDG}. However, in contrast to heavy quarkonia which have been studied comprehensively, 
heavy baryons have not been explored in much greater detail, both theoretically and 
experimentally, though the later can also provide similar information about the quark 
confinement mechanism as well as elucidating our knowledge about the nature of strong force 
by providing a clean probe of the interplay between the perturbative and the non-perturbative 
QCD. Experimentally only a handful of singly charmed baryons have been discovered \cite{PDG}. 
Experimentally the discovery of doubly charm baryon is controversial \cite{PDG}, whereas no 
triply heavy baryon has been observed yet. Moreover most of the observed charmed baryons do 
not have assigned quantum numbers yet. However it is expected that the large statistical 
sample that will be collected in experiments at BES-III, the LHCb, and the planned PANDA 
experiment at GS/FAIR may provide significant information for baryons with heavy quarks. In 
light of these existing and future experimental prospects on charm baryon studies, it is 
desirable to have model independent predictions from first principles calculations, such as 
from lattice QCD. Results from such calculations can be compared with those obtained from 
potential models which have been very successful in the case of charmonia and will naturally 
provide crucial inputs to the future experimental discovery. Various lattice QCD calculations 
were performed to compute only the ground states of various charm hadrons with spin up to 
${3\over 2}$, including quenched \cite{Lewis:2001iz, Mathur:2002ce} as well as full QCD [4-9].
Here we present a comprehensive study of the triply and doubly-charmed baryon spectra with 
spin up to ${7\over 2}$ for both parities. In addition we also discuss the quark mass 
dependence of various energy splittings among them due to hyperfine interactions as well as 
for spin-orbit coupling. The details work on triply charmed spectra was reported in Ref. 
\cite{Padmanath:2013zfa}.

\vspace{-0.2in}
\section{Numerical details}

\vspace{-0.4cm}
In recent years the Hadron Spectrum Collaboration (HSC) has exploited a dynamical anisotropic 
lattice formulation to extract highly excited hadron spectra.  Adopting a large anisotropy 
co-efficient $\xi=a_s/a_t=3.5$, with  $a_t~m_c\ll1$, it ensures that the standard relativistic 
formulation of fermions can also be used for charm quarks. Along with Symmanzik-improved gauge 
action the $N_f=$2+1 flavours fermionic fields were described using an anisotropic 
Shekholeslami-Wohlert action with tree-level tadpole improvement and stout-smeared spatial 
links. The temporal lattice spacing, $a_t^{-1}=5.67$GeV, was determined by equating the 
$m_{\Omega}$ to its physical value, resulting in a lattice spatial extension of 1.9 fm, which 
should be sufficiently large for a study of triply-charmed baryons. More details of the 
formulation of actions as well as the techniques used to determine the anisotropy parameters 
can be found in Refs. \cite{Edwards:2008ja, Lin:2008pr}. The lattice action parameters of 
the gauge field ensembles used in this work are given in \tbn{lattices}.

\bet[h]
\centering
\betb{cccc|cc|cc}
\hline
Lattice size & $a_t m_\ell$   & $a_t m_s$ & $N_{\mathrm{cfgs}}$ &  $m_\pi/$MeV & $a_t m_\Omega$ & $N_{\mathrm{tsrcs}}$ & $N_{\mathrm{vecs}}$\\\hline
$16^3 \times 128$ & $-0.0840$ & $-0.0743$ & 96 & 396        & $0.2951(22)$   & 4 & 64   
\\\hline
\eetb
\caption{Details of the gauge-field ensembles used. $N_{\mathrm{cfgs}}$  is the number of gauge-field 
configurations while $N_{\mathrm{tsrcs}}$ and $N_{\mathrm{vecs}}$ are the number of time sources per 
configuration and the number of eigenvectors used for each time source in the distillation method, 
respectively.}
\eet{lattices}

\vspace{-0.1cm}
\subsection{Operator construction and spin identification}
\vspace{-0.1cm}
\noindent Lattice computations of hadron masses proceed through the calculations of the Euclidean two 
point correlation functions, between creation operators at time $t_i$ and annihilation operators at 
time $t_f$.
\vspace{-0.2cm}
\beq
C_{ij}(t_f-t_i) = \langle 0|O_j(t_f)\bar{O}_i(t_i)|0\rangle = \sum_{n} {Z^{n*}_iZ^n_j\over 2 m_n} e^{-m_n(t_f-t_i)}  
\eeq{2pt} 

\vspace{-0.2cm}

\noindent{}The RHS is the spectral decomposition of such two point functions where the sum is over a discrete 
set of states. $Z^{n} = \langle 0|O_i^{\dagger}|n\rangle$ is the vacuum state matrix element, also called as 
overlap factor. We use a large basis of operators, constructed employing derivative-based operator construction 
formalism \cite{Edwards:2011jj}, including non-local operators constructed using up to two derivatives by 
which we are able to realize states up to spin $J=7/2$ for both the parities. Further the quark fields in these 
operators were distilled so as to compute the matrix of correlation functions with reduced contamination from 
the UV modes in the low energy physics that we are interested in \cite{Peardon:2009gh}. The continuum states get 
subduced \cite{Edwards:2011jj} over the irreps of the symmetry of the lattice ($\mathcal{O}_h$). For each of 
these irreps, we compute $N\times N$ matrix of correlation functions, where N is the number of operators used 
in each irrep as tabulated in the \tbn{operators}(a). Here $g$ and $u$ represents positive and negative parity. 
\tbn{operators}(b) gives the details of the subset of operators that are formed just by considering only the upper 
two components of the four component Dirac-spinors. This subset of operators are called non-relativistic as they 
form the whole set of creation operators in a leading order velocity expansion. 

\begin{table}
\small
\vspace*{-1.0cm}
\hspace{-0.6cm}
\parbox{.45\linewidth}{
\centering
\betb{ c | c  c  c  c  c  c }
\hline
          &\multicolumn{2}{c}{$G_1$}&\multicolumn{2}{c}{H}&\multicolumn{2}{c}{$G_2$} \\ \cline{2-7}
          &   $g$    &     $u$      &    $g$   &    $u$   &     $g$   &    $u$       \\ \hline
$ccc$     &    20    &      20      &     33   &     33   &      12   &     12       \\
$ccc_h$   &     4    &       4      &      5   &      5   &       1   &      1       \\
$ccc_{nr}$&     4    &       1      &      8   &      1   &       3   &      0       \\ \hline
$ccq$     &    55    &      55      &     90   &     90   &      35   &     35       \\
$ccq_h$   &    12    &      12      &     16   &     16   &       4   &      4       \\
$ccq_{nr}$&    11    &       3      &     19   &      4   &       8   &      1       \\
\hline
\multicolumn{7}{c}{(a)} 
\eetb
\label{nooperators}
}
\hspace{-0.4cm}
\parbox{.45\linewidth}{
\betb{c | c c  c  c | c c c c } 
\multicolumn{9}{c}{Non-Relativistic : SU(6)$\otimes$O(3)} \\ \hline
& \multicolumn{4}{c}{$ccc$} \vline & \multicolumn{4}{c}{$ccq$}   \\ \hline
\backslashbox{D}{J}  & $1/2$ & $3/2$ & $5/2$ & $7/2$ & $1/2$ & $3/2$ & $5/2$ & $7/2$ \\ \hline
0                    &   0   &   1   &   0   &   0   &   1   &   1   &   0   &   0   \\ \hline
1                    &   1   &   1   &   0   &   0   &   3   &   3   &   1   &   0   \\ \hline
2$_h$                &   1   &   1   &   0   &   0   &   3   &   3   &   1   &   0   \\ \hline
2                    &   2   &   3   &   2   &   1   &   6   &   8   &   5   &   2   \\ \hline
\multicolumn{9}{c}{(b)}
\eetb
\label{nroperators}
}
\vspace{-0.5cm}
\caption{(a) Total number of operators used for $ccc$ and $ccq$ baryons in each lattice irrep. 
(b) The description of the non-relativistic operators used. D stands for the number of derivatives, 
while J represents the continuum spin. In both $h$ and $nr$ stands for the hybrid and non-relativistic 
operators respectively.}
\vspace{-0.3cm}
\eet{operators}


We employ a variational method \cite{Dudek:2007wv} to extract the spectrum of baryon states from the matrix of correlation 
functions calculated by using the large basis of interpolating operators discussed in the previous 
subsection. The method proceeds by solving a generalized eigenvalue problem of the form 
\vspace{-0.3cm}
\beq
\nonumber C_{ij}(t)v_j^{(n)}(t,t_0) = \lambda^{(n)}(t,t_0)C_{ij}(t_0)v_j^{(n)}(t,t_0),
\eeq{a10}
where the eigenvalues, $\lambda^{(n)}(t,t_0)$ form the principal correlators and the eigenvectors are related 
to the overlap factors as $Z_i^{(n)} = \langle0|O_i|n\rangle = \sqrt{2E_n}\exp^{E_nt_0/2}v_j^{(n)\dagger}C_{ji}(t_0)$.
The energies are determined by fitting the principal correlators, while the spin identification of the states
are made by using these overlap factors as discussed in ref.\cite{Dudek:2007wv}. 

\bef[t]
\vspace*{-0.7cm}
\small
\parbox{.45\linewidth}{
\centering
  \includegraphics[width=70mm,trim=0 0 0 0 mm, clip=true]{3o2etc_splitting_spectrum_mev.eps}\\
(a)}
\hspace{0.75cm}
\parbox{.45\linewidth}{ 
\centering
  \includegraphics[width=70mm,trim=0 0 0 1 mm, clip=true]{3o2jpsi_ground_splitting_compare_HSC.eps}\\
(b)}
\caption{(a) Spin identified spectra of triply-charmed baryons with respect to $\frac32 m_{\eta_c}$ mass. 
The boxes with thick borders corresponds to the states with strong overlap with hybrid operators. 
The states inside the pink ellipses are those with relatively large overlap to non-relativistic 
operators. (b) Mass splitting of the ground state of $J^P=\frac32^+ ~\Omega_{ccc}$ from $\frac32$
times the mass of $J/\psi$ meson is compared for various lattice calculations. }
\eef{ccc_spectrum}

\vspace{-0.5cm}
\section{Results}

\vspace{-0.4cm}
In \fgn{ccc_spectrum}(a) we show the spin identified spectra of the triply charmed baryons where $3/2$ times 
the mass of $\eta_c$ is subtracted to account for the difference in the charm quark content \cite{Padmanath:2013zfa}. 
It is preferable to compare the energy splittings between the states, as it reduces the systematic uncertainty in the 
determination of the charm quark mass parameter in the lattice action and to lessen the effect of ambiguity 
in the scale setting procedure. Boxes with thicker borders correspond to those with a greater overlap onto 
the operators that are proportional to the field strength tensor, which might 
consequently be hybrid states \cite{Dudek:2012ag}. The states inside the pink ellipses have relatively large overlap with 
non-relativistic operators and should thus be well described in a quark model. One remarkable feature that 
one can observe is the number of states in the non-relativistic band exactly agree with expectations 
shown in \tbn{operators}(b). This agreement between the spectra and the expectations based on a model with the 
non-relativistic quark spins provides a clear signature of SU(6)$\otimes$O(3) symmetry in the spectra.

\bef[b]
\small
\parbox{.45\linewidth}{
\centering
  \includegraphics[width=70mm,trim=0 1 0 1 mm, clip=true]{ccs_etac_split.eps}\\
(a)}
\hspace{0.75cm}
\parbox{.45\linewidth}{ 
\centering
  \includegraphics[width=70mm,trim=0 1 0 1 mm, clip=true]{ccu_etac_split.eps}\\
(b)}
\caption{ Spin identified spectra of (a) $\Omega_{cc}$ and (b) $\Xi_{cc}$ baryon for both parities 
and with spins up to $\frac72$. The keys are same as in Figure 1(a).}
\eef{ccq_spectrum}

To assess the effect of radiative corrections on the co-efficient of the improvement term in the charm quark action 
 which could lead to significant change in the physical predictions, a second calculation was carried out after 
the co-efficient was boosted from the tree-level $c_s=1.35$ to $c_s=2.0$. As was seen for the $1^{--}$ and 
other excited states in the charmonium study \cite{Liu:2012ze}, we observe a shift in the energy difference, 
$m_{\Omega_{ccc}}-3/2m_{\eta_c}$, to be approximately 45 MeV \cite{Padmanath:2013zfa}. In \fgn{ccc_spectrum}(b) we compare our 
results for $m_{\Omega_{ccc}}-3/2m_{J/\psi}$ with other lattice calculations which use different 
discretization and so have distinct artefacts. Our results are consistent other results.

\fgn{ccq_spectrum} shows the spin identified spectra of the doubly charmed baryons. Here the spectra
is shown with the mass of $\eta_c$ subtracted from them so as to account for the difference in 
the charm quark content. The boxes and the pink ellipses represent the similar quantities as in \fgn{ccc_spectrum}.
Here again one can see the agreement between the number of states in the lower non-relativistic bands 
and the expectations as shown in the \tbn{operators}(b) providing a clear signature of SU(6)$\otimes$O(3) 
symmetry in the doubly charm baryon spectra also.

\begin{wrapfigure}{r}{80mm}
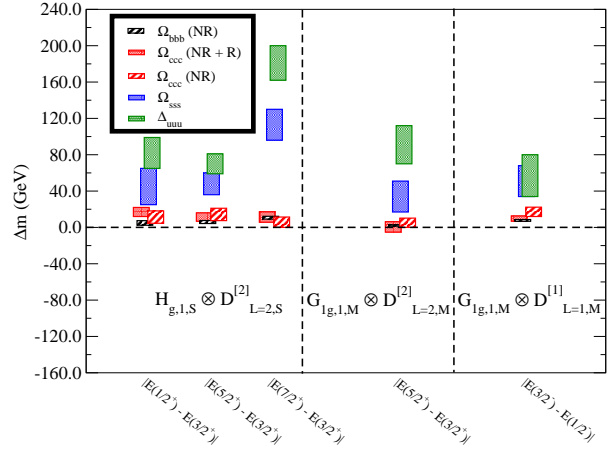

\begin{center}
  \incfig{0.32}{smsplittings.eps}\\
\caption{ Energy splittings between states with same L and S values, starting from light to heavy 
triple-flavoured baryons. For $\Omega_{bbb}$, results are with only non-relativistic operators 
\cite{Meinel:2012qz}. For $\Omega_{ccc}$, results from relativistic and non-relativistic as well as 
only non-relativistic operators are shown, and for the light and strange baryons results are with 
relativistic and non-relativistic operators \cite{Edwards:2011jj}.}
\end{center}
\label{ccc_socoup}
\vspace*{-0.1in}
\end{wrapfigure}

The spin dependent energy splittings provide important information about the nature of the interactions 
between different excitations. The most notable baryon energy splitting are the splittings due to 
spin-orbit coupling and the hyperfine splittings between spin-$\frac32^+$ and $\frac12^+$ states. In Figure 
\ref{ccc_socoup} we plot the absolute values of energy differences between energy levels which originate 
from the spin-orbit interaction of the following (L, S) pairs : (2,3/2-in the left), (2,1/2-in 
the middle) and (1,1/2-in the right column). We plot these splittings at varying quark masses 
from {\it light} to {\it bottom}. We identified these (L,S) pairs by finding the operators which 
incorporate these pairs and which have major overlaps to these states. For {\it bottom} baryons we used 
the data from Ref. \cite{Meinel:2012qz}, and for {\it light} and {\it strange} baryon results from Ref. 
\cite{Edwards:2011jj} are used. As one can see the degeneracy between these states is more or less satisfied 
for charm baryons \cite{Padmanath:2013zfa} as is also observed in {\it bottom} baryons. However, data with higher statistics 
is necessary to identify conclusively the breaking of this degeneracy at {\it charm} quark. 

For doubly-charmed baryons, in \fgn{ccq_ground}, we plot the hyperfine splittings between spin- 3/2 and 1/2 
along with other lattice estimations as well as with various other potential model calculations [19-24]. 
It is to be noted that our results for $\Xi_{cc}$ are at pion mass 396 MeV. However results from Ref. 
\cite{Alexandrou:2012xk, Briceno:2012wt, Namekawa:2013vu, Bali:2012ua} 
are extrapolated to the physical pion mass, while the Ref. \cite{Basak:2012py} results are at pion mass 390 
MeV and 340 MeV (black circle).
\vspace*{-0.2in}
\bef[t]
\small
\vspace*{-0.3in}
\parbox{.45\linewidth}{
\centering
  \incfig{0.55}{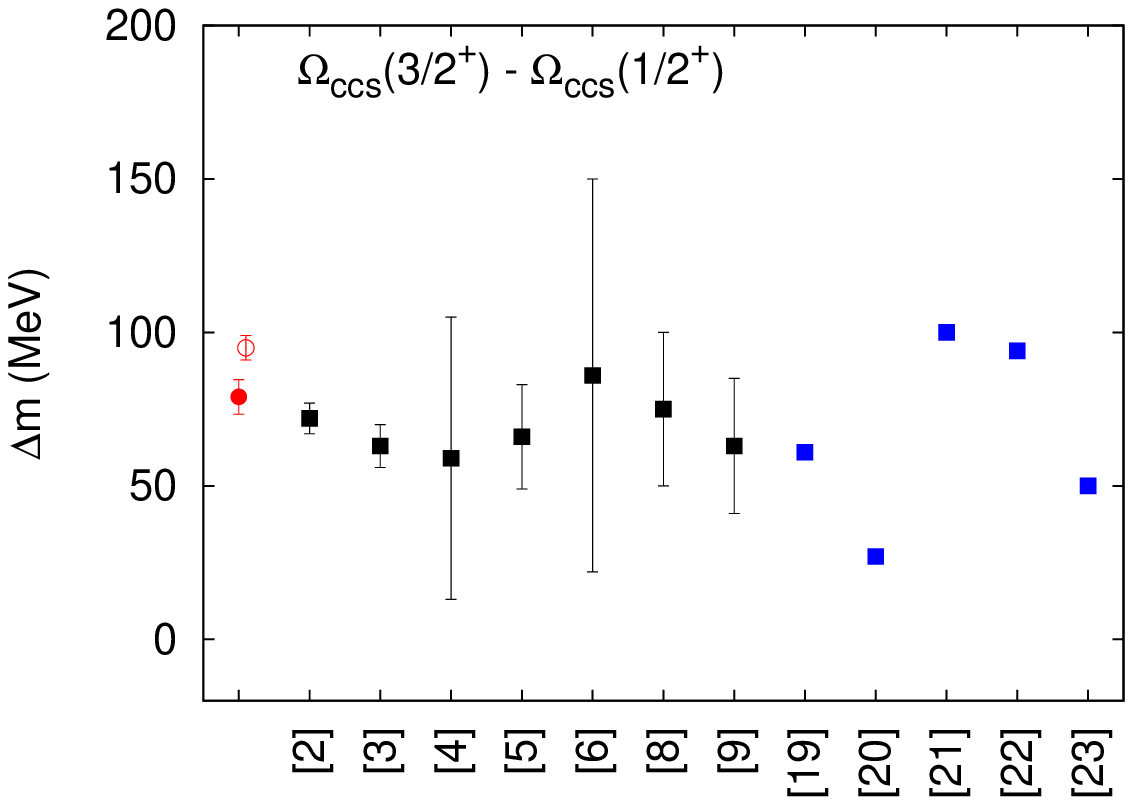}
}
\hspace{0.75cm}
\parbox{.45\linewidth}{ 
\centering
  \incfig{0.55}{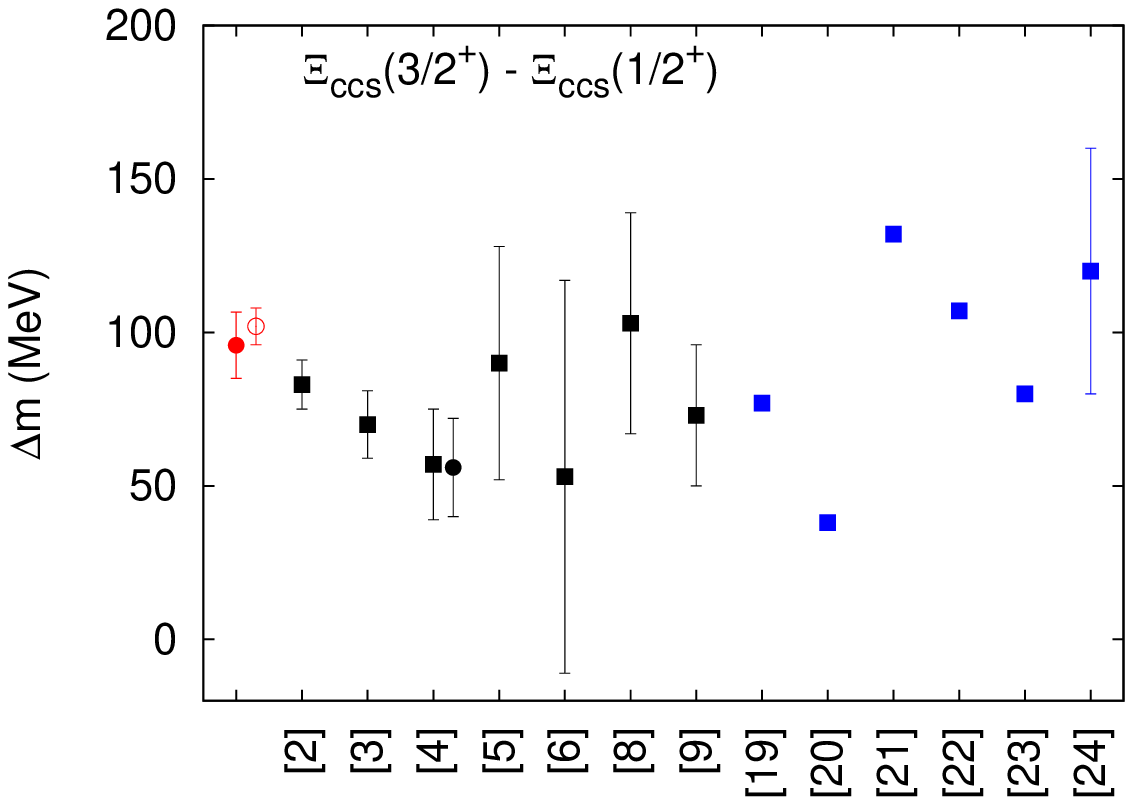}
}
\caption{Hyperfine splittings of $\Omega_{cc}$ and $\Xi_{cc}$ baryons are compared for various lattice and potential model calculations. The red circles are estimates from this work. 
The filled red circle is from $c_{sw}=1.35$, while the unfilled red circle is estimated from a boosted 
$c_{sw}=2.0$. The black squares represents other 
lattice results. The blue squares are estimates from various potential model 
calculations. }
\vspace*{-0.2in}
\eef{ccq_ground}

\bef[b]
\vspace*{-0.2in}
\small
\parbox{.45\linewidth}{
\centering
  \incfig{0.55}{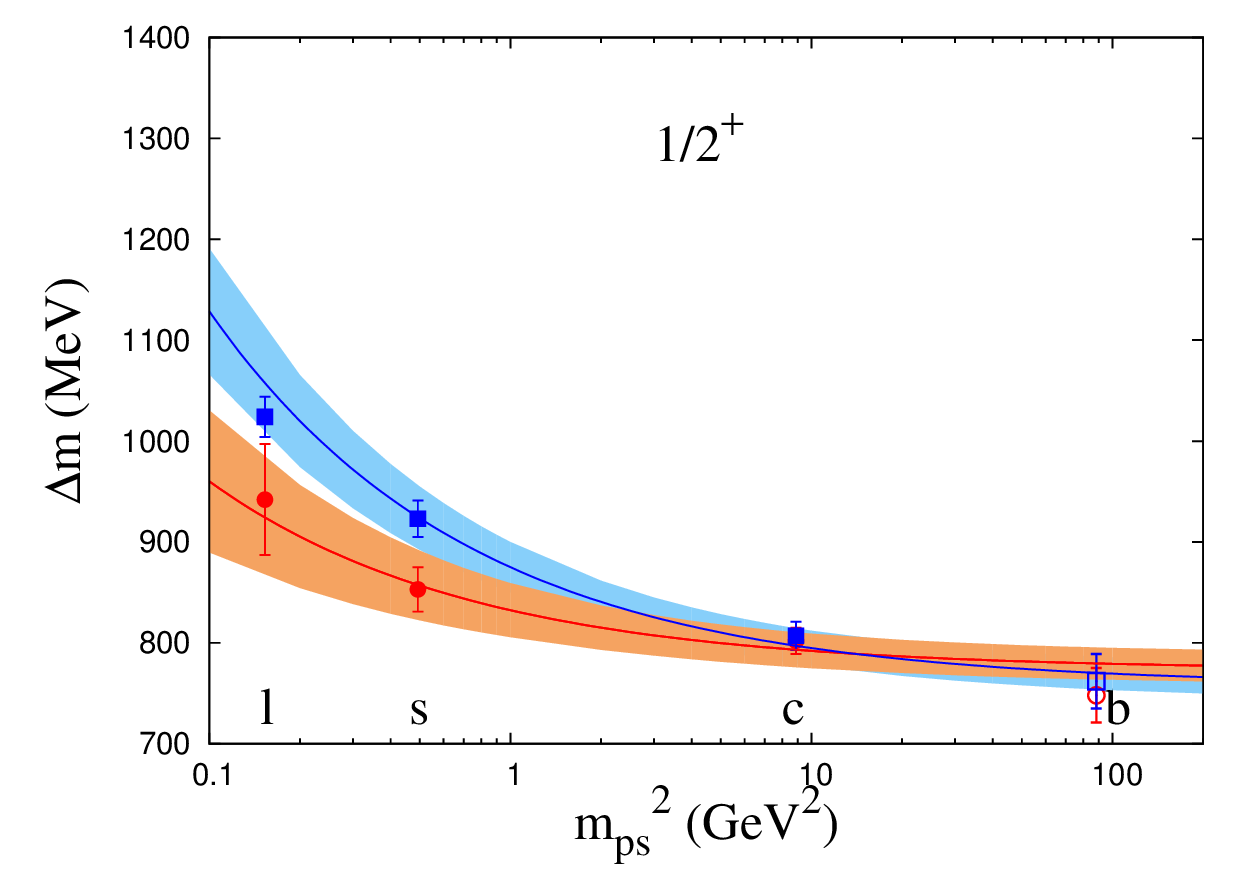}
}
\hspace{0.75cm}
\parbox{.45\linewidth}{ 
\centering
  \incfig{0.55}{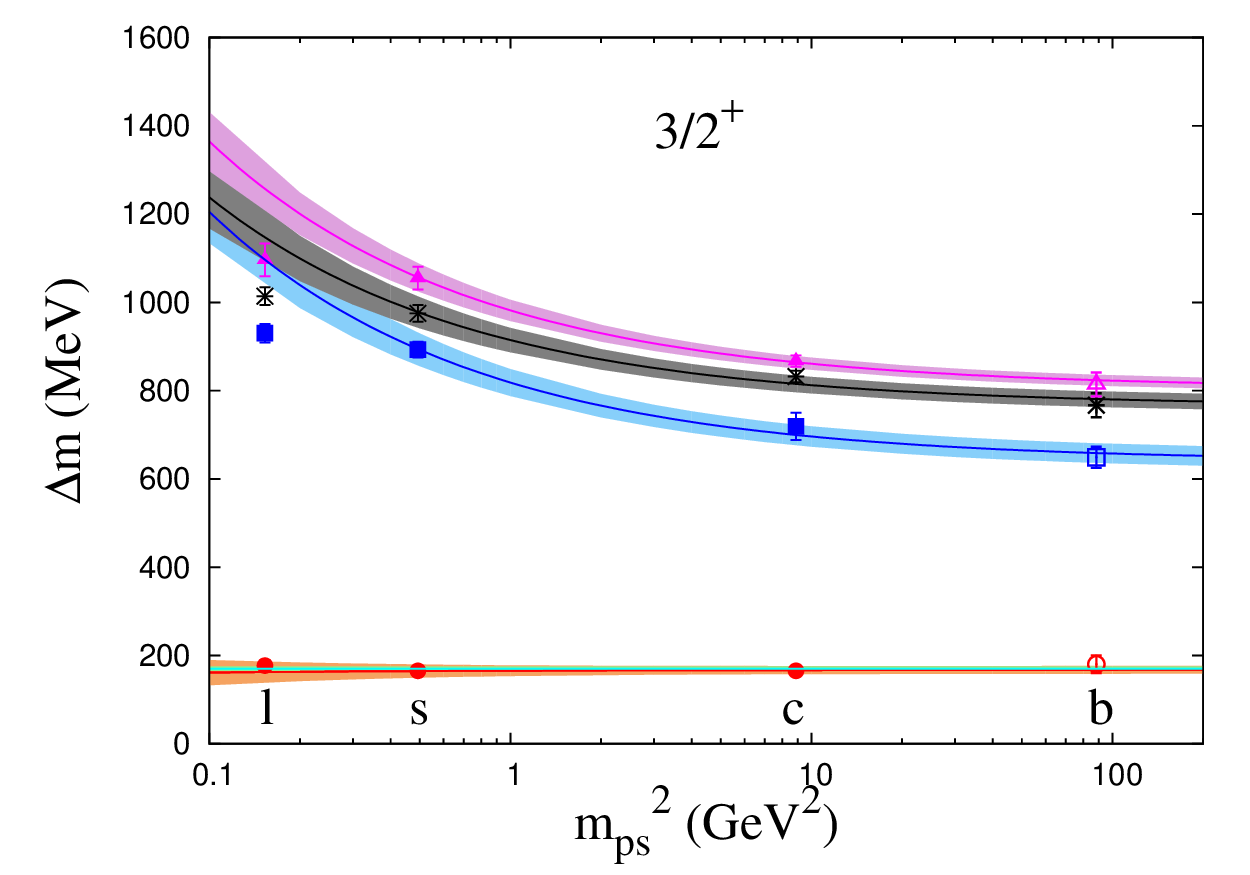}
}
\caption{Energy splittings of spin-$\frac12$ and $\frac32$ triple-flavored baryons from the isoscalar 
vector meson ground states are plotted against the square of the pseudo scalar meson masses. The keys for 
\textit{light}, \textit{strange} and \textit{bottom} data are the same as in Figure 2. Shaded 
regions are the extrapolations based on results from fitting the \textit{strange}, \textit{charm} 
and \textit{bottom} data to a heavy quark inspired form $\mathit{a+b/m_Q}$.}
\vspace*{-0.1in}
\eef{ccc_mq_dep}
\vspace*{0.2in}

In the heavy quark limit, naively one can expand the mass of a heavy hadron, with {\it n} heavy 
quarks as $m_{H_{nq}} = n~m_Q + A + B/m_Q + \mathcal{O}(1/m_{Q^2})$\cite{Jenkins:1996de}. Hence 
the energy splittings between the heavy hadrons can be expressed in the form $\mathit{a+b/m_Q}$.
In \fgn{ccc_mq_dep} we plot the energy splittings of spin-$\frac12$ and $\frac32$ triple-flavored 
baryons from the isoscalar vector meson ground states against the square of the 
pseudoscalar masses. Data for the {\it light} and {\it strange} are taken from Ref. \cite{Edwards:2011jj},
while the numbers for {\it bottom} quark are taken from Ref. \cite{Meinel:2012qz} which uses a
non-relativistic action. The shaded regions are the extrapolations based on results from fitting 
the data to the above form, excluding the {\it light} sector. 
One can observe that for various fits the fit estimates even pass through 
the {\it light} data even though they are not included in the fitting. We also make similar observations in other spin-parity channels. Study of
such energy splittings in the doubly {\it charm} sector allows us to make predictions in the \textit{bottom} hadrons. Fitting ($m_{\Xi^*_{cc}}-m_D$, $m_{\Omega^*_{cc}}-m_{D_s}$, $m_{\Omega_{ccc}}-m_{\eta_c}$) we 
get a prediction for $m_{\Omega_{ccc}} = 8050\pm10$ MeV, and fitting of ($m_{\Xi^*_{cc}}-m_{D^*}$, 
$m_{\Omega^*_{cc}}-m_{D_s^*}$, $m_{\Omega_{ccc}}-m_{J/\psi}$) allow us to predict $m_{B_c^*}-m_{B_c}=80\pm8$ MeV.

\vspace{-0.3cm}
\section{Conclusions}
\vspace{-0.3cm}
In this work we present the first calculation of the ground and excited state spectra of doubly and triply-charmed 
baryons by using dynamical lattice QCD. The spectra that we obtain have states with well-defined 
total spins up to 7/2 and the low lying states remarkably resemble the expectations of quantum 
numbers from SU(6) $\otimes$ O(3) symmetry. Various energy splittings including splittings due to 
hyperfine as well as spin-orbit coupling were studied and those are compared against the same energy 
splittings at other quark masses. From these energy splitting studies we also make predictions 
in the \textit{bottom} sector which are $m_{B_c^*}-m_{B_c} = 80\pm8$ MeV and 
$m_{\Omega_{ccb}} = 8050\pm10$ MeV. However, it is to be noted that we only mentioned statistical 
error in this work and the systematics from other sources like chiral extrapolation, lattice spacing are 
not addressed here. Also we have not incorporated multi-hadron operators which may effect some of the 
above conclusions, though to a lesser extent than their influence in the light hadron spectra.

\vspace{-0.3cm}
\section{Acknowledgements}
\vspace{-0.3cm}
We thank our colleagues within the Hadron Spectrum Collaboration.  Chroma~\cite{Edwards:2004sx} and
QUDA~\cite{Clark:2009wm,Babich:2010mu} were used to perform this work
on the Gaggle and Brood clusters of the Department of Theoretical
Physics, Tata Institute of Fundamental Research and at Lonsdale cluster
maintained by the Trinity Centre for 
High Performance Computing and at Jefferson
Laboratory. MP acknowledges support from the 
Trinity College Dublin Indian Research Collaboration Initiative and 
the CSIR, India for financial support through the SPMF.


\begin{thebibliography}{99}

\bibitem{PDG} The Review of Particle Physics : 
J. Beringer et al. (PDG), Phys. Rev. D86, 010001 (2012).

\bibitem{Lewis:2001iz} 
  R.~Lewis, N.~Mathur and R.~M.~Woloshyn,
  Phys.\ Rev.\ D {\bf 64}, 094509 (2001).

\bibitem{Mathur:2002ce} 
  N.~Mathur, R.~Lewis and R.~M.~Woloshyn,
  Phys.\ Rev.\ D {\bf 66}, 014502 (2002).

\bibitem{Basak:2012py} 
  S.~Basak, S.~Datta, M.~Padmanath, P.~Majumdar and N.~Mathur,
  PoS LATTICE {\bf 2012}, 141 (2012); \\ PoS LATTICE {\bf 2013}, {\it paper in preparation}.

\bibitem{Alexandrou:2012xk} 
  C.~Alexandrou {\it et al.},
  Phys.\ Rev.\ D {\bf 86}, 114501 (2012).


\bibitem{Briceno:2012wt} 
  R.~A.~Briceno, H.~-W.~Lin and D.~R.~Bolton,
  Phys.\ Rev.\ D {\bf 86}, 094504 (2012).

\bibitem{Durr:2012dw}
  S.~Durr, G.~Koutsou and T.~Lippert,
  Phys.\ Rev.\ D {\bf 86}, 114514 (2012)
  [arXiv:1208.6270 [hep-lat]].

\bibitem{Namekawa:2013vu} 
  Y.~Namekawa {\it et al.}  [PACS-CS Collaboration],
  arXiv:1301.4743 [hep-lat].

\bibitem{Bali:2012ua} 
  G.~Bali, S.~Collins and P.~Perez-Rubio,
  J.\ Phys.\ Conf.\ Ser.\  {\bf 426}, 012017 (2013).

\bibitem{Padmanath:2013zfa} 
  M.~Padmanath, R.~G.~Edwards, N.~Mathur and M.~Peardon,
  arXiv:1307.7022 [hep-lat].

\bibitem{Edwards:2008ja}
  R.~G.~Edwards, B.~Joo and H.~-W.~Lin,
  Phys.\ Rev.\ D {\bf 78}, 054501 (2008)
  [arXiv:0803.3960 [hep-lat]].

\bibitem{Lin:2008pr}
  H.~-W.~Lin {\it et al.}  [Hadron Spectrum Collaboration],
  Phys.\ Rev.\ D {\bf 79}, 034502 (2009)

\bibitem{Edwards:2011jj} 
  R.~G.~Edwards, J.~J.~Dudek, D.~G.~Richards and S.~J.~Wallace,
  Phys.\ Rev.\ D {\bf 84}, 074508 (2011).

\bibitem{Peardon:2009gh}
  M.~Peardon {\it et al.}  [Hadron Spectrum Collaboration],
  Phys.\ Rev.\ D {\bf 80}, 054506 (2009)

\bibitem{Dudek:2007wv} 
  J.~J.~Dudek, R.~G.~Edwards, N.~Mathur and D.~G.~Richards,
  Phys.\ Rev.\ D {\bf 77}, 034501 (2008).


\bibitem{Dudek:2012ag} 
  J.~J.~Dudek and R.~G.~Edwards,
  Phys.\ Rev.\ D {\bf 85}, 054016 (2012).

\bibitem{Liu:2012ze}
  L.~Liu {\it et al.}  [Hadron Spectrum Collaboration],
  JHEP {\bf 1207}, 126 (2012)
  [arXiv:1204.5425 [hep-ph]].

 
\bibitem{Meinel:2012qz} 
  S.~Meinel,
  Phys.\ Rev.\ D {\bf 85}, 114510 (2012).

\bibitem{Roberts:2007ni} 
  W.~Roberts and M.~Pervin,
  Int.\ J.\ Mod.\ Phys.\ A {\bf 23}, 2817 (2008).

\bibitem{Martynenko:2007je} 
  A.~P.~Martynenko,
  Phys.\ Lett.\ B {\bf 663}, 317 (2008).

\bibitem{Kiselev:2001fw}
  V.~V.~Kiselev {\it et al.},
  Phys.\ Usp.\  {\bf 45}, 455 (2002);

\bibitem{Ebert:2002ig} 
  D.~Ebert {\it et al.},
  Phys.\ Rev.\ D {\bf 66}, 014008 (2002).

\bibitem{Korner:1994nh} 
  J.~G.~Korner, M.~Kramer and D.~Pirjol,
  Prog.\ Part.\ Nucl.\ Phys.\  {\bf 33}, 787 (1994).

\bibitem{Brambilla:2005yk} 
  N.~Brambilla, A.~Vairo and T.~Rosch,
  Phys.\ Rev.\ D {\bf 72}, 034021 (2005).

\bibitem{Jenkins:1996de} 
  E.~E.~Jenkins,
  Phys.\ Rev.\ D {\bf 54}, 4515 (1996).

\bibitem{Edwards:2004sx} 
  R.~G.~Edwards {\it et al.},
  Nucl.\ Phys.\ Proc.\ Suppl.\  {\bf 140}, 832 (2005)
  [hep-lat/0409003].

\bibitem{Clark:2009wm} 
  M.~A.~Clark {\it et al},
  Comput.\ Phys.\ Commun.\  {\bf 181}, 1517 (2010).

\bibitem{Babich:2010mu} 
  R.~Babich, M.~A.~Clark and B.~Joo,
  arXiv:1011.0024 [hep-lat].


\end{thebibliography}
\end{document}